\documentclass[%
 amsmath,amssymb,
 aps,
]{revtex4-1}

\usepackage{graphicx}
\usepackage{dcolumn}
\usepackage{bm}
\usepackage{color}

\begin{document}

\title{ The Conformal Field Theory on the Horizon of BTZ Black Hole}
\author{Jingbo Wang}
\email{ shuijing@mail.bnu.edu.cn}
\affiliation{College of Physics and Electronic Engineering, Xinyang Normal University, Xinyang, 464000, P. R. China}
\author{Chao-Guang Huang}
\email{huangcg@ihep.ac.cn}
\affiliation{ Institute of High Energy Physics and Theoretical Physics Center for
Science Facilities, \\ Chinese Academy of Sciences, Beijing, 100049, People's Republic of China}
 \date{\today}
\begin{abstract}
In three dimensional spacetime with negative cosmology constant, the general relativity can be written as two copies of SO$(2,1)$ Chern-Simons theory. On a manifold with boundary the Chern-Simons theory induces a conformal field theory--WZW theory on the boundary. In this paper, it is show that with suitable boundary condition for BTZ black hole, the WZW theory can reduce to a massless scalar field on the horizon.
\end{abstract}
\maketitle
\section{Introduction}
In three dimensional spacetime, the general relativity become simplified since it has no local degrees of freedom \cite{carbook1}. Indeed the theory is equivalent to Chern-Simons theory with suitable gauge group \cite{at1,witten1}. It is a surprise that the black hole solution can exist when theory has negative cosmology constant $\Lambda<0$. This black hole-so called BTZ black hole \cite{btz1} can have an arbitrary high entropy which is difficult to understand since the theory has no local degrees of freedom.

This mystery can be understood if one starts from the  Chern-Simons formula. It is a standard result that on a manifold with boundary the Chern-Simons theory induces a Wess-Zumino-Witten (WZW) theory on the boundary which is a conformal field theory. Carlip use this WZW theory to explain the entropy of the BTZ black hole \cite{carlip1}. Later it was shown that, for the boundary at conformal infinity rather than the horizon, the Chern-Simons theory reduces to a Liouville theory on the boundary \cite{chp1,rs1}. This Liouville theory has the right central charge  to give the entropy of BTZ black hole if one use the Cardy formula \cite{cardy1,cardy2}. For a review along this line, see Ref.\cite{carlip2}.

There are other conformal field theories which start from Brown and Henneaux's seminal work \cite{bh1}. They observed that the asymptotic symmetry group of $AdS_3$ is generated by two copies of Virasoro algebra, which correspond to a conformal field theory. This result can be seen as pioneering work of $AdS_3/CFT_2$ \cite{kra1}. Based on this result, the entropy of BTZ black hole can be calculated \cite{str1,bss1}, which matches the Bekenstein-Hawking formula.

But most of those conformal field theories are taken to be at conformal infinity (although with exceptions, such as \cite{carlip1,mp1,mit1}). A physical more appealing location should be the horizon of black hole. In this paper, we consider the field theory just on the horizon. Starting from the Chern-Simons theory, with suitable boundary condition, it is shown that the WZW theory reduces to a chiral massless scalar field on the horizon. So on the BTZ horizon, there are two chiral massless scalar field since the 3D general relativity contain two copied of Chern-Simons theories.

The paper is organized as follows. In section II, we summary the relation between gravity, Chern-Simons theory and the WZW theory. In section III, the BTZ black hole is considered. With suitable boundary condition, the boundary WZW theory reduces to a chiral massless scalar field theory. Section IV is the conclusion.
\section{Gravity, Chern-Simons theory and WZW theory}
As first shown in Ref.\cite{at1}, $(2+1)-$dimensional general relativity can be written as a Chern-Simons theory. For the case of negative cosmology constant $\Lambda=-1/L^2$, one can define two SO$(2,1)$ connection 1-form
\begin{equation}\label{1}
    A^{(\pm)a}=\omega^a\pm \frac{1}{L} e^a,
\end{equation}where $e^a$ and $\omega^a$ are the co-triad and spin connection 1-form respectively. Then up to boundary term, the first order action of gravity can be rewritten as
\begin{equation}\label{2}\begin{split}
    I_{GR}[e,\omega]=\frac{1}{8\pi G}\int e^a \wedge ({\rm d}\omega_a+\frac{1}{2}\epsilon_{abc}\omega^b \wedge \omega^c)-\frac{1}{6L^2}\epsilon_{abc}e^a\wedge e^b \wedge e^c\\=I_{CS}[A^{(+)}]-I_{CS}[A^{(-)}],
\end{split}\end{equation}
where $A^{(\pm)}=A^{(\pm)a}T_a$ are SO$(2,1)$ gauge potential, and the Chern-Simons action is
\begin{equation}\label{3}
    I_{CS}[A]=\frac{k}{4\pi}\int Tr\{A\wedge {\rm d}A+\frac{2}{3}A\wedge A \wedge A\},
\end{equation}
with
\begin{equation}\label{4}
    k=\frac{L}{4G}.
\end{equation}
Similarly, the CS equation
\begin{equation}\label{5}
    F^\pm={\rm d}A^{(\pm)}+A^{(\pm)} \wedge A^{(\pm)}=0
\end{equation}is equivalent to the requirement that the connection is torsion-free and the metric has constant negative curvature.
The equation implies that the potential $A$ can be locally written as
\begin{equation}\label{5a}
    A=g^{-1}{\rm d}g.
\end{equation}

When the manifold has a boundary, a boundary term must be added. Assume the boundary has topology $\partial M=R\times S^1$. The usual boundary term is
\begin{equation}\label{6}
    I_{bd}=\frac{k}{4\pi}\int_{\partial M}Tr A_u A_{\tilde{u}},
\end{equation}where $u$ and $\tilde{u}$ are two coordinates on the boundary. The boundary condition is chosen to be
\begin{equation}\label{7}
    \delta A_u|_{\partial M}=0,\quad or \quad \delta A_{\tilde{u}}|_{\partial M}=0
\end{equation}
depend on the condition.

With the boundary term, the total action, $I_{CS}[A]+I_{bd}[A]$, is not gauge-invariant under the gauge transformation
\begin{equation}\label{8}
    \bar{A}=g^{-1}Ag+g^{-1}{\rm d}g.
\end{equation}To restore the gauge-invariant, the Wess-Zumino-Witten term is introduced for the first boundary condition\cite{ogu1,carlipbd1}:
\begin{equation}\label{9}
    I^+_{WZW}[g^{-1},A_u]=\frac{1}{4\pi}\int_{\partial M}Tr(g^{-1}\partial_u g g^{-1}\partial_{\tilde{u}}g+2g^{-1}\partial_{\tilde{u}}gA_u)+\frac{1}{12\pi}\int_M Tr(g^{-1}{\rm d}g)^3,
\end{equation}
which is chiral WZW action for a field $g$ coupled to a background gauge potential $A_u$.

With the WZW term, the full action is gauge-invariant
\begin{equation}\label{10}
    (I_{CS}+I_{bd})[\bar{A}]+kI^+_{WZW}[e^{-1},\bar{A}]=(I_{CS}+I_{bd})[A]+kI^+_{WZW}[g^{-1},A].
\end{equation}

Thus, the gauge transformation $g$ become dynamical at the boundary, and are described by the WZW action which is a conformal field theory. Those `would-be gauge degrees of freedom'\cite{carlipwb1} are present because the gauge invariant is broken at the boundary.
\section{The boundary action on the horizon of BTZ black hole}
In the previous section, the boundary of manifold can be arbitrary. If the horizon of BTZ black hole is considered, more reduction can be made due to the special property of the horizon.
\subsection{The BTZ black hole}
To study the physics at horizon, it is more suitable to use advanced Eddington coordinate. The metric of BTZ black hole can be written as
\begin{equation}\label{11}
    ds^2=-N^2 dv^2+2 dv dr+r^2 (d\varphi+N^\varphi dv)^2.
\end{equation}
Choose the following co-triads \cite{awd1}
\begin{equation}\label{12}
    l_a=-\frac{1}{2}N^2 dv+dr,\quad n_a=-dv,\quad m_a=r N^\varphi dv+r d\varphi,
\end{equation}
which gives the following connection:
\begin{equation}\label{13}
    A^{-(\pm)}=-(N^\varphi\mp \frac{1}{L})dr-\frac{N^2}{2} d(\varphi\pm\frac{v}{L}),\quad  A^{+(\pm)}=-d(\varphi\pm\frac{v}{L}),\quad A^{2(\pm)}=r (N^\varphi\pm\frac{1}{L})  d(\varphi\pm\frac{v}{L}),
\end{equation}
where $A^{\pm}=(A^0\pm A^1)/\sqrt{2}$.

Define new variables which are useful later,
\begin{equation}\label{14}
 u=\varphi-\frac{v}{L}, \quad  \tilde{u}=\varphi+\frac{v}{L}.
\end{equation}
A crucial property of the connection is that, on the whole manifold, one has
\begin{equation}\label{15}
    A^{(+)}_u\equiv 0,\quad A^{(-)}_{\tilde{u}}\equiv 0.
\end{equation}
Since the topology of the space-section is cylinder, which is non-trivial, the vacuum Chern-Simons equation $F=0$ will be solved by non-periodic group element \cite{rs1}
\begin{equation}\label{16}
    A=Q^{-1} {\rm d} Q.
\end{equation}
For a general SO$(2,1)$ group element $Q(\tilde{u},u,r)$, using the Gauss decomposition, it can be written as
\begin{equation}\label{17}
    Q=\left(
                  \begin{array}{cc}
                    1 & \frac{1}{\sqrt{2}}x_1 \\
                    0 & 1 \\
                  \end{array}
                \right)
                \left(
                         \begin{array}{cc}
                           e^{-\Psi_1/2} & 0 \\
                           0 & e^{-\Psi_1/2} \\
                         \end{array}
                       \right)
                       \left(
                         \begin{array}{cc}
                           1 & 0 \\
                           -\frac{1}{\sqrt{2}}y_1 & 1 \\
                         \end{array}
                       \right).
\end{equation}
Within this parameter, the WZW action is \cite{chp1}
\begin{equation}\label{18}\begin{split}
    kI_{WZW}=\frac{k}{4\pi}\int_{\partial M}du d\tilde{u}\frac{1}{2}(\partial_u \Psi \partial_{\tilde{u}} \Psi-e^\Psi (\partial_u x \partial_{\tilde{u}} y+\partial_u y \partial_{\tilde{u}} x)).
   \end{split}\end{equation}

\subsection{Gauge transformation}
Now we consider the gauge transformation (\ref{8}) with group element $g_1$ for the $A^{(+)}$. In following we omit the superscript ${(+)}$.

To preserve the boundary condition
\begin{equation}\label{19}
    \delta A_u|_{\partial M}=0,
\end{equation}the gauge transformation should be $g_1=g_1(r,\tilde{u})$.
But it is not enough. This boundary condition can't tell us whether we are dealing with a black hole or not, so more restricted boundary conditions are need. Near the horizon, a small parameter $\epsilon=r-r_+$ can be defined, and $N^2\approx 2 \kappa \epsilon$, thus
\begin{equation}\label{20}
    A^-_{\tilde{u}}\approx-\kappa \epsilon.
\end{equation}
Since this condition reflect the property of the horizon, we want the gauge transform to keep this property, thus
\begin{equation}\label{21}
    \bar{A}^-_{\tilde{u}}\sim O(\epsilon)=C_1 \epsilon.
\end{equation}

Assume the gauge transformation is given by SO$(2,1)$ group element
\begin{equation}\label{22}
    g_1(x_1,y_1,\Psi_1)=\left(
                  \begin{array}{cc}
                    1 & \frac{1}{\sqrt{2}}x_1 \\
                    0 & 1 \\
                  \end{array}
                \right)
                \left(
                         \begin{array}{cc}
                           e^{-\Psi_1/2} & 0 \\
                           0 & e^{-\Psi_1/2} \\
                         \end{array}
                       \right)
                       \left(
                         \begin{array}{cc}
                           1 & 0 \\
                           -\frac{1}{\sqrt{2}}y_1 & 1 \\
                         \end{array}
                       \right),
\end{equation}
under the gauge transformation (\ref{8}),
\begin{equation}\label{23}
  \bar{A}^-=e^{\Psi_1} (A^--A^2 x_1+A^+ x_1^2/2+dx_1),
\end{equation}
since $A^2, A^+$ are both finite at horizon, to keep the boundary condition (\ref{21}), one need
\begin{equation}\label{24}
    x_1(r,\tilde{u})=\epsilon h(\tilde{u}),
\end{equation}
where $h(\tilde{u})$ is a finite function at horizon. And also $\Psi_1(r,\tilde{u})$ is finite at horizon.

The other component transforms into
\begin{equation}\label{25}\begin{split}
    \bar{A}^2=A^2 (1-e^{\Psi_1} y_1 x_1)-A^+ x_1(1-e^{\Psi_1} y_1 x_1/2)+A^- e^{\Psi_1} y_1 +d \Psi_1+e^{\Psi_1} y_1d x_1,\\
    \bar{A}^+=A^+ e^{-\Psi_1}(1-e^{\Psi_1} y_1 x_1/2)^2+A^2 y_1(1-e^{\Psi_1} y_1 x_1/2)+A^- e^{\Psi_1} y_1^2/2+y_1 d\Psi_1+dy_1 +y_1^2e^{\Psi_1}dx_1/2.
\end{split}\end{equation}
Those components are required to be finite at the horizon, so gives
\begin{equation}\label{26}
    y_1(r_+)=finite, \quad \Psi_1(r=r_+)=finite.
\end{equation}
So the second term in the action (\ref{18}) vanish
\begin{equation}\label{27}
2 e^{\Psi_1} \partial_a x_1 \partial_b y_1 \backsim \epsilon= 0
\end{equation}
on the horizon. The final action on the horizon is
\begin{equation}\label{28}\begin{split}
    kI_{WZW}=\frac{k}{4\pi}\int_{\partial M}du d\tilde{u}\frac{1}{2}\partial_u \Psi_1 \partial_{\tilde{u}} \Psi_1\\
    =\frac{k}{4\pi L}\int_{\partial M}d\varphi dv [(\partial_v \Psi_1)^2- L^2 (\partial_{\varphi} \Psi_1)^2],
   \end{split}\end{equation}
   with $\Psi_1$ depend only on $\tilde{u}=\varphi+\frac{v}{L}$. So it is a chiral massless scalar field.

The similar results can be get for the $A^{(-)}$, which gives another chiral massless scalar field $\Psi_2$ depending only on $u$.
\section{Conclusion}
In this paper, the field theory on the horizon of BTZ black hole is investigated. Starting from the Chern-Simons formula, one get a chiral WZW theory on any boundary. Restrict to the horizon, this WZW theory reduces further to a chiral massless scalar field theory. Since the general relativity is equivalent to two copies of CS theory, the final theory on the horizon is two chiral massless scalar field theory with opposite chirality.

Compared with the conformal field theories on the conformal boundary, the massless scalar field theory-which is also a conformal field theory--is more revelent to black hole physics. It is just on the horizon. But the central charge of this theory is $c=1$ \cite{cft1}, which is too small to account the entropy of the BTZ black hole if one use the Cardy formula. 

The conformal symmetry here is different with that appears in Carlip's effective description of the black hole entropy in arbitrary dimension \cite{carlip4}. As noticed in \cite{carlip3}, the symmetry of this paper is on the $``\varphi-v$ cylinder", while the symmetry of \cite{carlip4} is on the $``r-v$ plane".

In the previous work \cite{wmz,wang1,wh1,wh2,wh3,hw1,hw2}, it was shown that the boundary degrees of freedom can also be described by a BF theory. Since both the BF theory and the massless scalar field theory are on the horizon, the relation between those two theories need further investigated.

\acknowledgments
 This work is supported by the NSFC (Grant No. 11690022 and No. 11647064).

\bibliography{cft2}
\end{document}